# Effect of Tamm surface states on hot electron generation and Landau damping in nanostructures metal-semiconductor

Alexander V. Uskov*[1], Igor V. Smetanin[1], Igor E. Protsenko[1], Morten Willatzen,[2] and Nikolay V. Nikonorov[3]

[1] P. N. Lebedev Physical Institute, Leninsky pr. 53, 119991 Moscow, Russia

[2] Beijing Institute of Nanoenergy and Nanosystems, Chinese Academy of Sciences, Beijing, 100083, PR China

[3] ITMO University, Kronverksky Pr. 49, St. Petersburg 197101, Russia

**ABSTRACT:** The hot electron generation in plasmonic nanoparticles is the key to efficient plasmonic photocatalysis. In the paper, we study theoretically for the first time the effect of Tamm states (TSs) at the interface metal-semiconductor on hot electron generation and Landau damping (LD) in metal nanoparticles. TSs can lead to resonant hot electron generation and to the LD rate enhanced by several times. The resonant hot electron generation is reinforced by the *transition* absorption due to the jump of the permittivity at the metal-semiconductor interface. Since electron states in the metal and the quasi-discrete TS are coupled *coherently* ("bound state in continuum"), the absorption spectrum of light by electrons has a Fano-type shape. Our results demonstrate clearly the importance of taking into account details of the semiconductor band structure and surface states at the metal-semiconductor interface, including Tamm surface states, for a proper description of the hot carrier generation and LD. The results are in correspondence with earlier experimental works on coherent electron transport and chemical-induced damping in plasmonic nanostructures. Thus, by judicious selection of semiconductor materials with Tamm surface states one can engineer decay rates and hot carrier production for important applications, such as photodetection and photochemistry.

## 1. INTRODUCTION

Hot electrons, generated with nanoplasmonic structures, find many applications [1], in particular, in the energy conversion technologies, such as photocatalysis (so called "hot electron chemistry") [2-4]. In metal nanoparticles of small sizes, the hot electron generation is dominated by photon absorption of electrons in the metal during their collisions with the nanoparticle boundary (the interface between metal and its surrounding). The hot electron generation manifests itself by virtue of the size-dependent broadening of plasmonic resonance in metal nanoparticles. Although this kind of plasmonic broadening is well-known, ("1/R−law", "Kreibig broadening", often called also as "Landau broadening" or "Landau damping (LD)"; see [5] and references therein), it continues to attract significant attention of both theoreticians and experimentalists [6-21], since Landau damping (we will use this term) affects the characteristics of nanoplasmonic devices – see [22]. Hot electron generation and LD depend strongly on characteristics of the interface metal-surrounding [2-4, 6,7, 16-21]. In particular, electron states of adsorbates at the surface of the metal nanoparticle can strongly affect the decay of plasmons in the nanoparticles, an effect known as the so-called "chemical interface damping" (CID) [6, 16, 19,20]. Correspondingly, the measurement of LD rates provides an effective tool both in the physics of nanostructure surface [6-7, 16, 18, 20] and in studies of hot electron generation for applications in photocatalysis [2-4, 16-21]..

In Ref. [10], a quantum-mechanical method has been developed so as to calculate the Landau damping rate in plasmonic nanoparticles with complex interface "metal-surrounding", including the complicated behavior of the electron potential, the electron mass, and the dielectric constant at the boundary of the nanostructure. Recently [21], LD in metal nanoparticles covered with semiconductor shell (so-called, "hybrid plasmonics structures" [18]) has been evaluated for the first time using this approach, and it was shown that hot electron generation in hybrid structures can be substantially enhanced mainly due to the effect of the quasi-discrete electron state in the conduction band of the semiconductor shell.

In this work, we present the first theoretical study of the effect of Tamm states (TSs) [23,24] at the interface metal-semiconductor for hot electron generation and LD in metal nanoparticles. We show that Tamm States, located in the semiconductor bandgap, can lead to the resonant hot electron generation and to an enhanced LD rate. The resonance hot generation is strengthened by the *transition* absorption [25] due to the jump of the permittivity at the metal-semiconductor interface. Since electron states in the metal and quasi-discrete TS are coupled *coherently* (where the coupling can be viewed as a "bound state in the continuum" [26] with "coherent transport" [27] between them), the single-electron absorption spectrum of light has a Fano-type shape.

Note that in [28] we have studied enhanced electron photoemission *from* metal to surrounding via Tamm State, the state relating to a bandgap in metal. In contrast, in present paper we study hot electron generation *inside* metal nanoparticle due to resonant absorption with participation of Tamm state, and this TS is based on the semiconductor band gap which is more typical for material systems. This hot electron generation in metal nanoparticle leads to Landau damping of plasmons in the nanoparticle, which is characterized by calculated Kreibig coefficient.

In Sect. 2, the formulation of the problem is given, and the model to calculate Landau damping is presented briefly. In Sect. 3, we describe the origin of Tamm surface quasi-levels at the interface between the metal and the semiconductor, basing on a simplified Kronig-Penney model. In Sect. 4, the probability for a *single* electron in the metal to absorb a photon during its collision with the interface is calculated. In Sect. 5, we calculate the rate of hot electron generation and the Kreibig coefficient for the Landau damping. Sect. 6 concludes our findings.

## 2. FORMULATION OF PROBLEM AND THE MODEL TO CALCULATE LANDAU DAMPING

The model that captures all the key features of the hot carrier excitation is shown in Fig.1. An electromagnetic wave of frequency $\omega$ illuminates a metal nanoparticle with permittivity $\varepsilon_m$, buried into semiconductor medium with permittivity $\varepsilon_s$, and excites the localized surface plasmon (LSP) in it (see Fig.1a). An electron in the metal collides with the nanoparticle boundary and absorbs a photon of energy $\hbar\omega$ during the collision and becomes hot. This photon absorption leads to broadening of LSP resonance ("Landau damping") and can be described by adding the term $\Delta\gamma_{LD}$ to the damping rate $\gamma_c$ of electrons in bulk metal [5]: $\Delta\gamma_{LD} = A\mathrm{v}_F / L_{nano}$, where $\mathrm{v}_F$ is the Fermi velocity of the metal, $L_{nano}$ is the characteristic size of the nanoparticle, commensurate with the volume-to-surface ratio. According to [10], the Kreibig coefficient $A$ is given as

$$A = A_{mat} A_{geom} \qquad (1)$$

where $A_{geom}$ is determined by the nanoparticle size and shape, and the spatial distribution of the electric field in the nanoparticle (see also [10, 29]). For example, in the quasistatic approximation for the dipole mode in spherical nanoparticle, $A_{geom} = 1$ [29]. On the other hand, $A_{mat}$ depends on both the frequency and material parameters and, in particular, on the characteristics of the interface between the metal nanoparticle and the surroundings [10]:

$$A_{mat} = 0.5 \cdot \left(\omega^3 / \omega_p^2\right) \cdot \left(\hbar / \varepsilon_o \mathrm{v}_F\right) \cdot K_R \qquad (2)$$

where $\omega_p$ is the plasma frequency of the metal (we assume the Drude approximation, $\varepsilon_m = 1 - \omega_p^2 / \left[\omega(\omega + i\gamma_c)\right]$), and $\varepsilon_o$ is the permittivity of vacuum. The coefficient $K_R$ determines the photon absorption rate $R$ per unit square of nanoparticle surface [1/(m$^2$·s)]: $R = K_R \cdot |F_m|^2$, where $F_m$ is the component of the electric field inside the metal nanoparticle, *normal* to its interface. In [10] (see also [29]), the coefficient $A_{mat}$ was calculated assuming an *infinite* potential barrier at the metal interface. On the other hand, in [21] $A_{mat}$ was obtained for more complicated hybrid plasmonic nanostructures where the infinite-barrier model assumption does not apply. In the present paper, $A_{mat}$ is determined when the Tamm Surface state is positioned at the metal-semiconductor interface.

Note that the key feature of the [10] is the use of a "locally flat surface" approximation valid for smooth nanoparticles in which $L_{nano}$, and thus the radius-of-curvature $R_{cur}$, are much larger than the de Broglie electron wavelength $ƛ$ (~several angstroms). Thus, the "locally flat surface approximation" is valid for nanoparticles larger than a few nm and it allows reduction of the complicated quantum-mechanical problem involving full quantization of states to a simple one-dimensional problem treated here – see below.

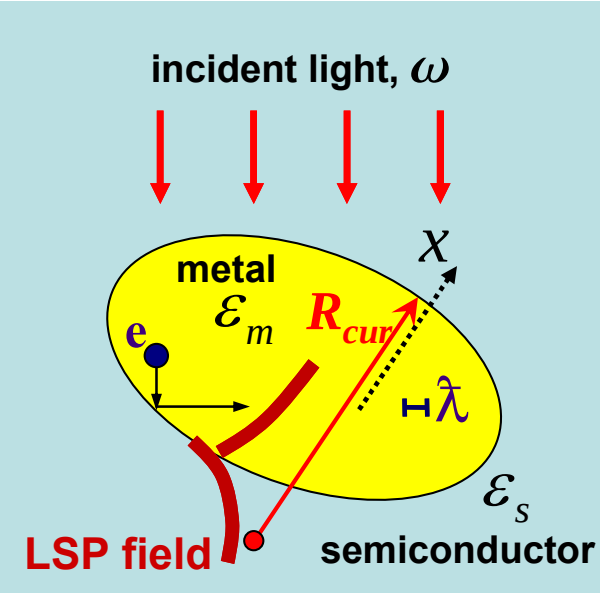

(a)

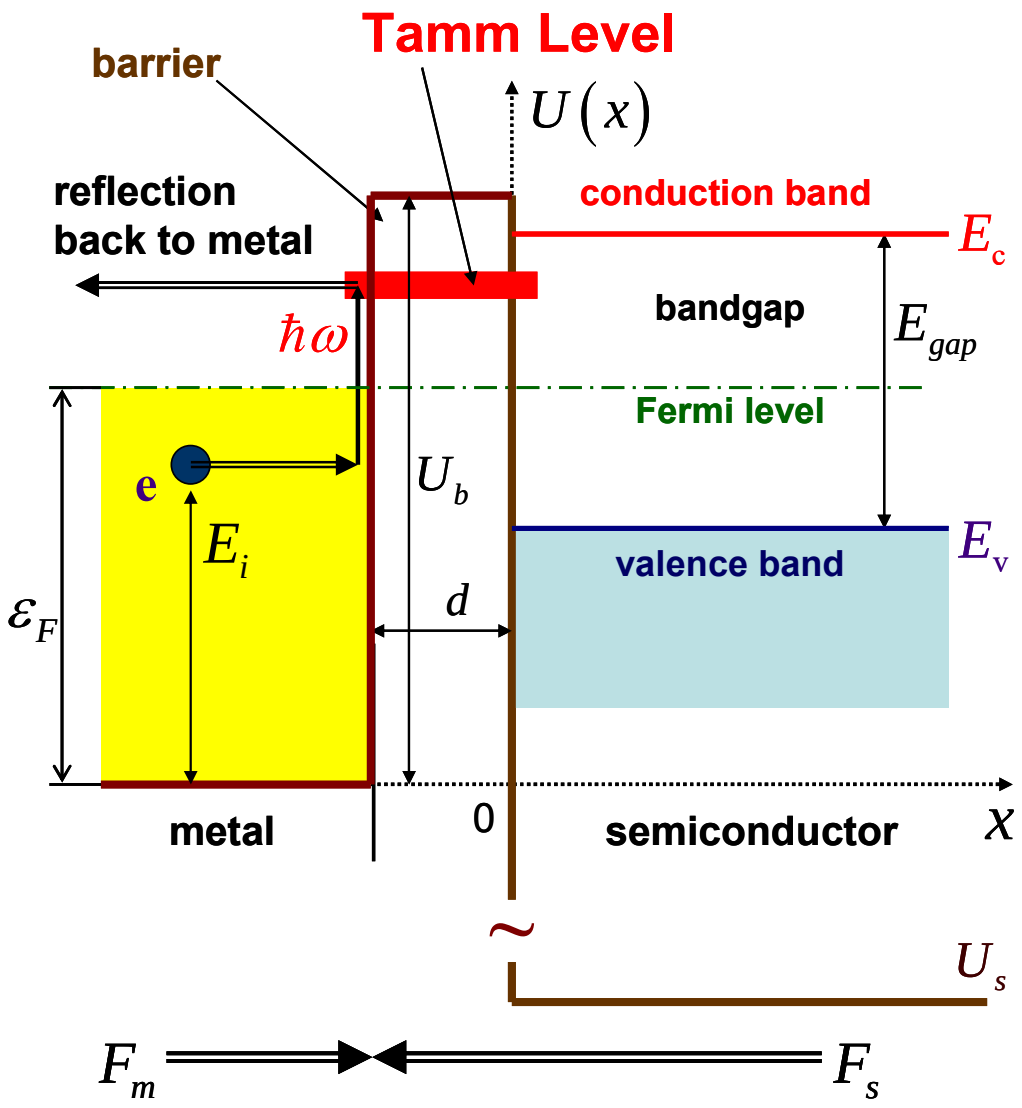

(b)

**Figure 1.** (a) A metal nanoparticle (yellow) with dielectric constant $\varepsilon_m$, buried into a dielectric/semiconductor material (blue) with dielectric constant $\varepsilon_s$, is illuminated by light of frequency $\omega$, exciting the localized surface plasmon (LSP) mode. The red arrow illustrates the local radius-of-curvature $R_{cur}$ which is much larger than the de Broglie electron wavelength $ƛ$. For smooth nanoparticles, $R_{cur}$ is of the order of the nanoparticle size $L_{nano}$ $\left(R_{cur} \sim L_{nano}\right)$. Bold red lines illustrate a LSP mode excited in the nanoparticle. (b) Energy diagram of the metal-semiconductor structure in the direction $x$ normal to the surface of the structure (Fig. 1a). For metals, the conduction band is shown with the Fermi energy $\varepsilon_F$ (=5.5eV) measured from the bottom of this band. For the semiconductor, one forbidden band (the bandgap) as well as the conduction and valence bands generated by a 1D periodic potential [see Eq.(3)] are shown; $E_c$ is the bottom of the conduction band, and $E_v$ is the top of valence band. A thin barrier of thickness $d$ and height $U_b$ is shown between the metal and semiconductor. A Tamm surface quasi-discrete energy level, located in the semiconductor bandgap, is shown as a red bar. $F_s$ and $F_m$ are the amplitudes of the field in the semiconductor and metal, respectively, and $\varepsilon_s F_s = \varepsilon_m F_m$. An electron in the metal (dark blue circle) with initial energy $E_i$ collides with the metal-semiconductor interface and absorbs the photon energy $\hbar\omega$ before it is reflected back to the metal.

## 3. TAMM SURFACE QUASI-DISCRETE STATE AT THE METAL-SEMICONDUCTOR INTERFACE

The "metal-semiconductor" structure is modeled by a 1D electron potential $U(x)$ in the direction $x$ normal to the surface of the structure (see Fig. 1a) ("locally flat surface" approximation):

$$U(x)=\begin{cases} 0, & x<-d \quad \text{metal} \\ U_b>0, & -d<x<0 \quad \text{barrier} \\ U_s+\alpha\cdot\sum_{n=1}^{n=\infty}\delta(x-b-a\cdot n), & x>0 \quad \text{semiconductor} \end{cases} \qquad (3)$$

The potential inside the metal is taken as a constant (as in Sommerfeld's model), $U(x<-d)=0$, which is the bottom of the conduction band in the metal – see Fig.1b. The Fermi energy in the metal, measured from the bottom of the metal conduction band, is $\varepsilon_F$. The semiconductor in the potential is modeled by a 1D periodic potential of $\delta$ – functions similar to Ref.[23] by Igor Tamm; $a$ is the period of the potential or the lattice constant in semiconductor. This periodic potential generates the band structure in the semiconductor, depicted partly in Fig.1b. We assume that a thin barrier of the thickness $d$ and the height $U_b$ is sandwiched between the metal and the 1D periodic potential. The barrier can represent, for instance, the effect of coupling together a real metal and semiconductor lattices. If the barrier thickness $d$ is infinite ($d=\infty$), the existence of a *discrete* energy level (Tamm level) inside the forbidden band (bandgap) of the semiconductor and below the barrier height $U_b$ [23-24] is possible. In this quantum state, Tamm State, electron motion is finite in the direction *x*, since electrons are reflected both from the barrier and the semiconductor whereby the electron is "trapped". If the thickness $d$ is finite ($d<\infty$), electrons can escape from the TS by tunneling through the barrier into the metal. As a result, the Tamm energy level for $d<\infty$ *becomes broadened* [24], so that one should talk on a Tamm quasi-discrete energy level and the continuous electron energy spectrum in the structure in whole. In this case, one can talk also on a "bound state in continuum" [26] that leads to a Fano-type absorption spectrum – see below. The width of the Tamm state is affected strongly by the barrier characteristics.

Note that the band structure of the 1D crystal representing the semiconductor band structure, is determined completely by the dimensionless parameter $S=m\alpha a/\hbar^2$ [$m$ is the free electron mass, $\alpha$ is the coefficient in the front of $\delta$ -functions in Eq. (3)], the lattice constant *a* and the energy level $U_s$. Below, we choose *A*, *a* and $U_s$ by such manner so as to have the bandgap width $E_{gap}=3\text{eV}$, and to locate the Fermi level exactly in the middle of the bandgap: $S=6$, $a=5.1\,\text{Å}$ and $U_s=-8.9\,\text{eV}$. The metal Fermi energy is $\varepsilon_F=5.5\text{eV}$. Note that in Fig.1b, the forth bandgap above the energy $U_s$ is shown (see also the calculation of the band structure in Supplementary Material SI-1). In calculations, the barrier height $U_b$ is chosen to be equal to the energy of the conduction band bottom $E_c$ (that is, $U_b=E_c$).

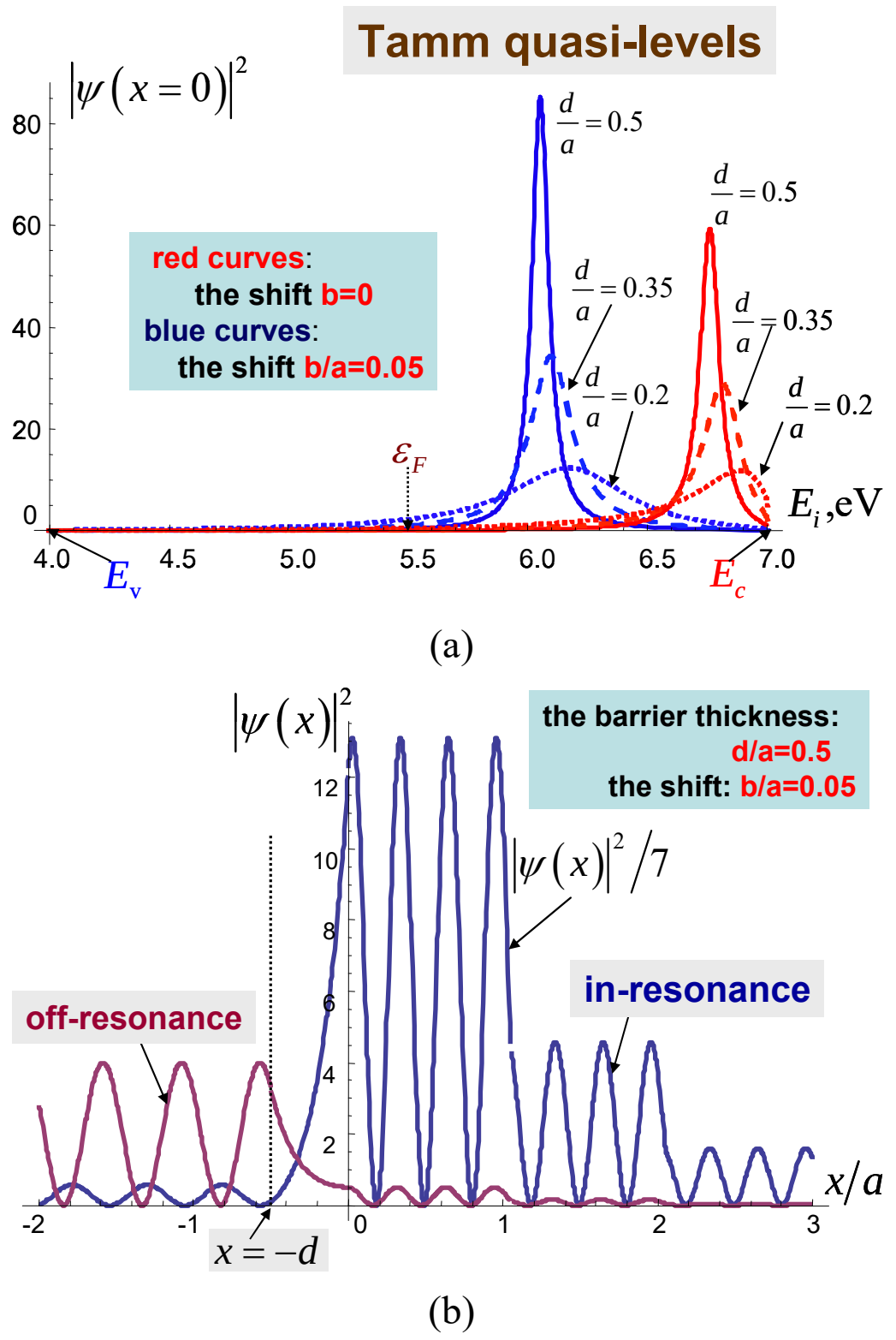

**Figure 2**. (a) $|\psi(x=0)|^2$ as function of the incident electron energy $E_i$ at the semiconductor boundary $x=0$ for various barrier thicknesses *d* (solid curves: $d=0.5a$, dashed curves: $d=0.35a$, dotted curves: $d=0.2a$). The resonances illustrate Tamm quasi-discrete energy states. Red curves: the shift $b=0$; blue curves: the shift $b/a=0.05$ (b) Spatial behavior of the electron wavefunction when the electron is in resonance with the Tamm surface state (blue), and when the electron is off-resonance (violet).

Solving the Schrödinger equation with the electron potential $U(x)$ (see Supplementary Material, SI-3), one finds the wavefunction $\psi(x,E_i)$ of an electron with energy $E_i$ which is incident normally from the metal to the semiconductor. Fig.2a shows the square module of the wavefunction $\psi(x,E_i)$ at $x=0$ (i.e., at the boundary between the barrier and the semiconductor) as a function of the energy $E_i$, for energies inside the bandgap of the semiconductor, i.e., between the bottom of the conduction band $E_c$ and the top of the valence band $E_v$ (see Fig.1b). For simplicity, in the calculations, the effective electron mass in the metal and the barrier are assumed to be equal to the free-electron mass $m$. The maxima in Fig.2a demonstrate clearly Tamm quasi-levels inside the bandgap: When $E_i$ approaches the Tamm level, the electron resonantly penetrates into the Tamm State at the interface between metal and semiconductor (see also Fig. 2b). We stress that with increasing *d* the position of the Tamm quasi-level tends exactly to the position of *discrete* Tamm levels at $d=\infty$. As seen from Fig.2a, the Tamm level energy is sensitive to the value of the shift *b* of the periodic potential relatively the barrier: Increasing the value of *b* (shifting of the periodic potential in the positive direction of the axis *x*) leads to a red shift of the resonance. The width of the Tamm quasi-level (and also its location) depends strongly on the barrier thickness *d*: Decreasing *d* results in broadening of the resonances. Such behavior is understandable: Decreasing of the barrier thickness *d* leads to shortening of the electron lifetime at the Tamm level and, correspondingly, to broadening of the level due to faster tunneling of the electron to the metal.

Fig. 2b shows the spatial behavior of the wavefunction $|\psi(x)|^2$ in and off resonance. In resonance (blue curve), $|\psi(x)|^2$ *increases exponentially* inside the barrier and reaches large values compared to $|\psi(x)|^2$ off resonance (violet), where the wavefunction *decays exponentially* inside the barrier. In calculations, the barrier thickness $d=0.5a$, and the shift $b=0.05a$. Note that namely this resonance behavior of the wavefunction leads to the resonant absorption of the photon energy.

## 4. ABSORPTION OF LIGHT BY ELECTRON OF METAL COLLIDING WITH METAL-SEMICONDUCTOR INTERFACE

If an electric field of optical frequency $\omega$ exists in the metal-semiconductor structure, the field has a component normal to the metal-semiconductor interface, then a *single* electron, incident from metal to the interface metal-semiconductor, can reflect back to the metal with absorption of a photon $\hbar\omega$ from the field – see Fig. 1a and 1b. The probability $p_a$ of this process can be calculated using time-dependent perturbation theory for the continuous spectrum – see [10, 30-31] and Supporting Information in [21] where, in the latter reference, a detailed derivation shows,

$$p_a = p_a\left(\hbar\omega, E_{i,x}, E_{i,\parallel}\right) = \left(\mathrm{v}_{f,x}/\mathrm{v}_{i,x}\right)\cdot\left|C_-\right|^2 \qquad (4)$$

where $E_{i,x}$ ($E_{i,\parallel}$) is the kinetic energy of the electron motion in the metal normal (parallel) to the interface; $\mathrm{v}_{i,x}$ ($\mathrm{v}_{f,x}$) is the initial (the final) velocity of the electron in the metal along axis *x*; and the amplitude $C_-$ is calculated as

$$C_- = \frac{ie}{\omega}\frac{F_m}{m\mathrm{v}_{fz}}\cdot\left\{\int_{-\infty}^{+\infty} dx\frac{d\psi_i}{dx}\psi_f + \right.$$
$$\left. +\left(\frac{\varepsilon_m}{\varepsilon_s}-1\right)\left[\int_{-d}^{+\infty} dx\frac{d\psi_i}{dx}\psi_f + \frac{1}{2}\cdot\psi_f\left(x=-d\right)\cdot\psi_i\left(x=-d\right)\right]\right\} \qquad (5)$$

Correspondingly, the probability (4) can be written as

$$p_a = c_a\left(\hbar\omega, E_{i,x}, E_{i,\parallel}\right)\cdot\left|F_m\right|^2 \qquad (6)$$

where $c_a\left(\hbar\omega, E_{i,z}, E_{i,\parallel}\right) = \left(\mathrm{v}_{f,z}/\mathrm{v}_{i,z}\right)\cdot\left|C_-\right|^2/\left|F_m\right|^2$. In Eq. (4), $\psi_i$ and $\psi_f$ are the electron wavefunctions in the initial state, with the energies $E_{i,z}$ and $E_{i,\parallel}$, and in the final state with the energies $E_{f,z} = E_{i,z} + \hbar\omega$ and $E_{f,\parallel} = E_{i,\parallel} = E_\parallel$, respectively. One should stress that since the electron mass does not change along the axis *x*, the probability $p_a$ does not depend on the energy $E_{i,\parallel}$: $c_a = c_a\left(\hbar\omega, E_{i,x}\right)$.

The first term in braces in Eq.(5) (referred as term I below) describes the absorption of a photon (LSP) with energy $\hbar\omega$ due to collision of the electron with the potential barrier [see Eq. (3) and Fig. 1b]. If $U(z)\equiv 0$ (the barrier is absent), the term I is equal to zero, and the second term (term II below), proportional to $\left[\left(\varepsilon_m/\varepsilon_s\right)-1\right]$, describes the pure *transition absorption* due to the jump of the dielectric constant at the interface [25, 31]. Of course, in real structures with a finite barrier at the interface and with $\varepsilon_s \neq \varepsilon_m$, the complex amplitudes in Eq.(5) *interfere* with each other, and adding of the mechanisms of photon absorption together becomes nontrivial. Obviously, the role of this term II is enhanced with decreasing $\varepsilon_s$ in Eq.(5) – see also below.

Note that the above approach to calculate the absorption probability $p_a$ has been used to evaluate the probability of the electron *photoemission* in the *surface* photoeffect [30-33].

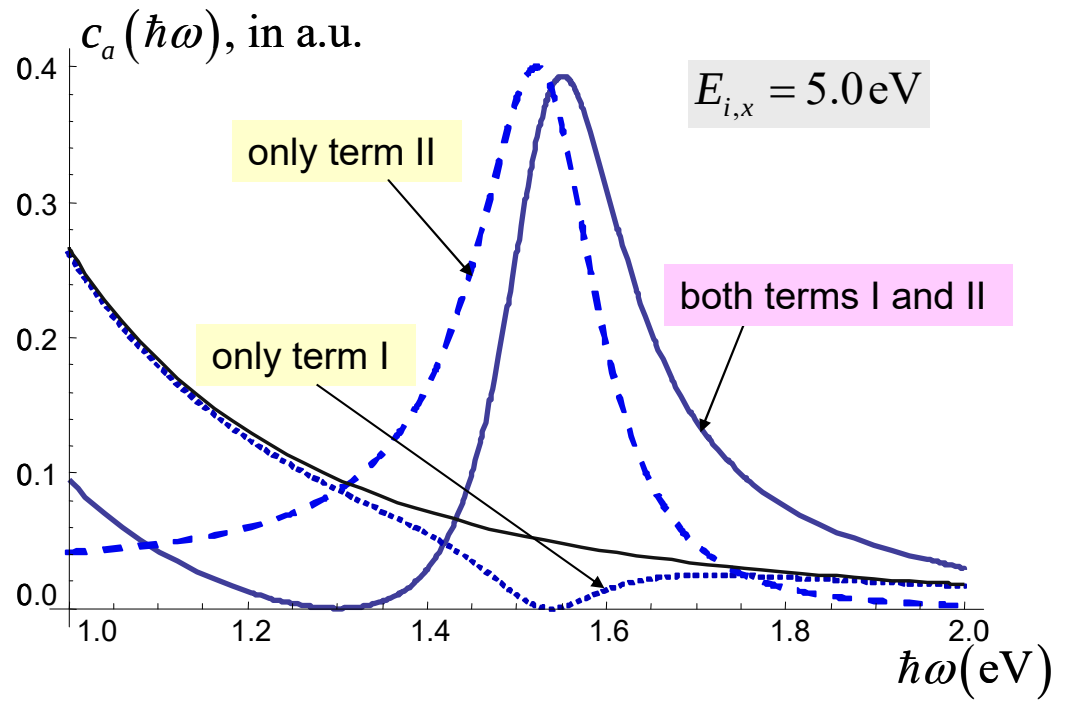

(a)

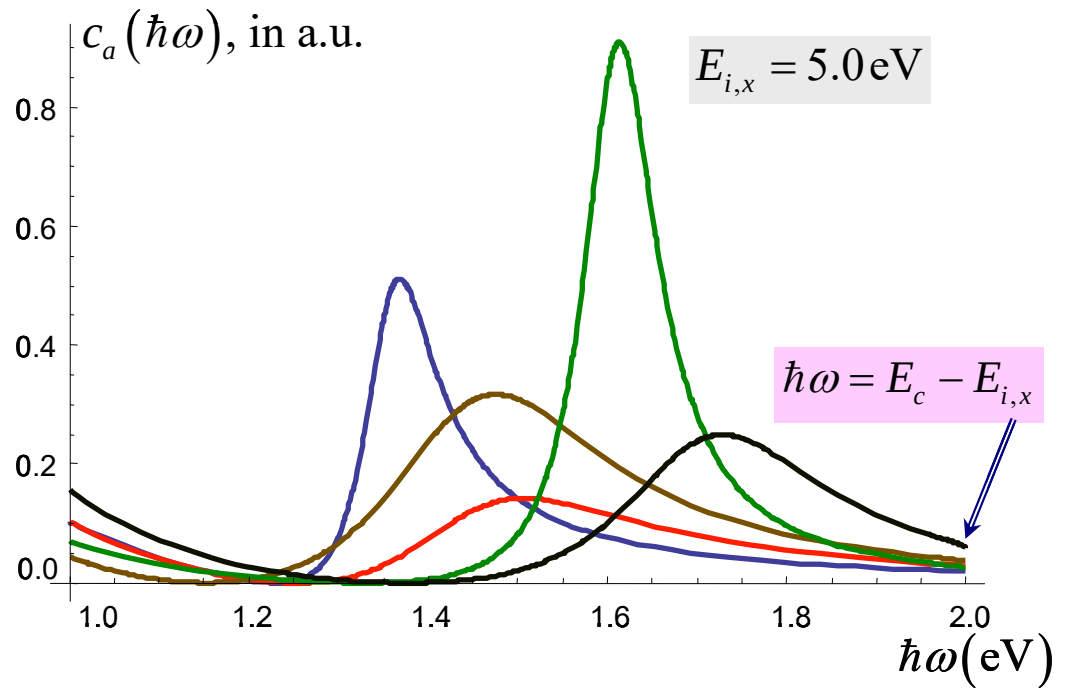

(b)

**Figure 3**. Spectrum of the probability coefficient $c_a = c_a\left(\hbar\omega, E_{i,x}\right)$ for $E_{i,x}$ =5eV. (a) Black curve is for the infinite barrier at the metal interface ($U_b=\infty$); solid blue curve: both terms I and II in braces in Eq.(5) are taken into account; dotted (dashed) blue curves: only term I (II) are effective; $d/a$ =0.4, $b/a=0.02$, $\varepsilon_s$ =3. (b) The blue curve is $d/a$ =0.5, $b/a$ =0.03, $\varepsilon_s$ =5; the red curve is $d/a$ =0.3, $b/a$ =0.03, $\varepsilon_s$ =5; the brown curve is $d/a$ =0.3, $b/a$ =0.03, $\varepsilon_s$ =3; the green curve is $d/a$ =0.5, $b/a$ =0.012, $\varepsilon_s$ =2. The blue arrow shows the photon energy

$\hbar\omega = E_c - E_{i,x}$ when the electron in its final state reaches the top of the bandgap.

Fig. 3 shows the spectrum of the probability coefficient $c_a = c_a(\hbar\omega, E_{i,x})$ at a given initial electron energy $E_{i,x}$=5eV for various parameters of the interface: the barrier thickness $d$, the shift $b$ and the semiconductor permittivity $\varepsilon_s$. In Fig.3a, the black curve is for the infinite barrier at the interface, $U_b = \infty$ (as in [10, 29]), and is considered as a reference. The solid blue curve takes into account both terms I and II in braces in Eq.(5); the dotted (dashed) curve is calculated only with term I (term II). Here and below we use the dielectric function of the metal $\varepsilon_m$ with $\hbar\omega_p$ =9eV, $\hbar\gamma_c$ =0.07eV.

A few important observations should be made. First of all, one notes that the resonances in the curves are manifestations of the Tamm surface quasi-levels inside the bandgap – see Fig.1 and 2.

Furthermore, all the resonances are Fano-like (compare, in particular, the solid curve with the reference black curve in Fig. 3a) as a consequence of the "bound state in the continuum" effect (see [26] and references therein). One can say that the continuum states in the metal and the Tamm surface state are coupled to each other *coherently,* since the states are described with a *common* wavefunction over the whole structure, i.e., in the metal and the semiconductor. Thus, "coherent electron transfer" occurs between metal and semiconductor [27].

Finally, a comparison of curves in Fig. 3a shows clearly that the contribution of the term II, which depends on the permittivity ratio $\varepsilon_m/\varepsilon_d$, dominates in braces in Eq.(5), leading to a strong dependence of the hot carrier generation and the LD on the dielectric constant $\varepsilon_s$, similarly to the situation for hybrid plasmonic structures [21], and a decrease of $\varepsilon_s$ leads to enhancement of the coefficient $c_a$ – see Fig. 3b. Fig. 3b shows that the resonance width depends strongly on the barrier thickness *d*, and the resonance location is affected by the shift *b* as in Fig. 2a.

## 4. RATE OF HOT ELECTRON GENERATION AND KREIBIG COEFFICIENT

To find the absorption coefficient $K_R$ in Eq. (2) for the coefficient $A_{mat}$, one must sum over all electrons of the metal, colliding with the interface of the metal [10]:

$$K_R(\hbar\omega) = \int_{k_{i,x}>0} \frac{2d\boldsymbol{k}_i}{(2\pi)^3} f_F(\boldsymbol{k}_i)\cdot\left[1 - f_F(\boldsymbol{k}_f)\right]\cdot \mathrm{v}_{i,x}\cdot c_a(\hbar\omega, E_{i,x}) \quad (7)$$

where $\boldsymbol{k}_i$ ($\boldsymbol{k}_f$) is the wave vector of electron in the initial (final) state in metal; $\mathrm{v}_{i,x} = \hbar k_{i,x}/m$ is the initial electron velocity, normal to the interface; $f_F(\mathbf{k}_{i(f)})$ is the Fermi distribution. Since the probability coefficient $c_a$ depends only on the energy $E_{i,x} = \hbar^2 k_{i,x}^2/2m$, the formula (7) can be rewritten as

$$K_R(\hbar\omega) = \int_0^{+\infty} dE_{i,x}\cdot D(E_{i,x},\hbar\omega)\, c_a(\hbar\omega, E_{i,x}) \quad (8)$$

where

$$D(E_{i,x},\hbar\omega) = \frac{m k_B T_e}{2\pi^2\hbar^3}\cdot$$
$$\cdot\ln\left[\left(1+\exp\left(\frac{\varepsilon_F - E_{i,x}}{k_B T_e}\right)\right)\Big/\left(1+\exp\left(\frac{\varepsilon_F - E_{i,x} - \hbar\omega}{k_B T_e}\right)\right)\right] \quad (9)$$

is the number of electrons in metal at initial energy $E_{i,x}$ which absorb a photon $\hbar\omega$. $T_e$ is the electron temperature in metal. Note that in Eq.(9) we assumed that $\hbar\omega >> k_B T_e$. The rate of generation of hot electrons with the energy $E_{f,x} = E_{i,x} + \hbar\omega$ is thus

$$G_{hot}(E_{f,x},\hbar\omega) = D(E_{f,x} - \hbar\omega, \hbar\omega)\, c_a(\hbar\omega, E_{i,x} - \hbar\omega) \quad (10)$$

If the lifetime of hot carriers does not depend on $E_{f,x}$, their energy distribution is proportional to $G_{hot}(E_{f,x},\hbar\omega)$. Below we give results for the rate distribution $G_{hot}(E_{f,x},\hbar\omega)$, and for the coefficient $A_{mat}$ calculated according Eqs. (8-10) and (2), respectively.

Fig. 4 shows the distribution $G_{hot}(E_{f,x})$ of the hot electron generation rate over the energy $E_{f,x}$ for a given photon energy $\hbar\omega$ =1.3eV for a structure with parameters $d/a$ =0.4, $b/a$ =0.04, $\varepsilon_s$ =2 (red curve) together with $G_{hot}(E_{f,x})$ for the infinite barrier at the metal interface ($U_b = \infty$) – black curve. One can see the peak in the blue curve due to resonant optical transitions of electrons in the metal to the Tamm state near the top of the bandgap $E_c$ substantially exceeds the distribution $G_{hot}(E_{f,x})$ for the case of an infinite barrier (black curve). At the red-side wing of the Fano-type resonance, the generation rate is suppressed and is below the rate for the infinite barrier. Nevertheless, the generation rate, integrated over the energy $E_{f,x}$, for this structure is larger than the integrated generation rate for the case of an infinite barrier – see below for the results of the coefficient $A_{mat}$.

The green curve in Fig. 4 illustrates the initial electron distribution $D(E_{i,x},\hbar\omega)$ in the metal.

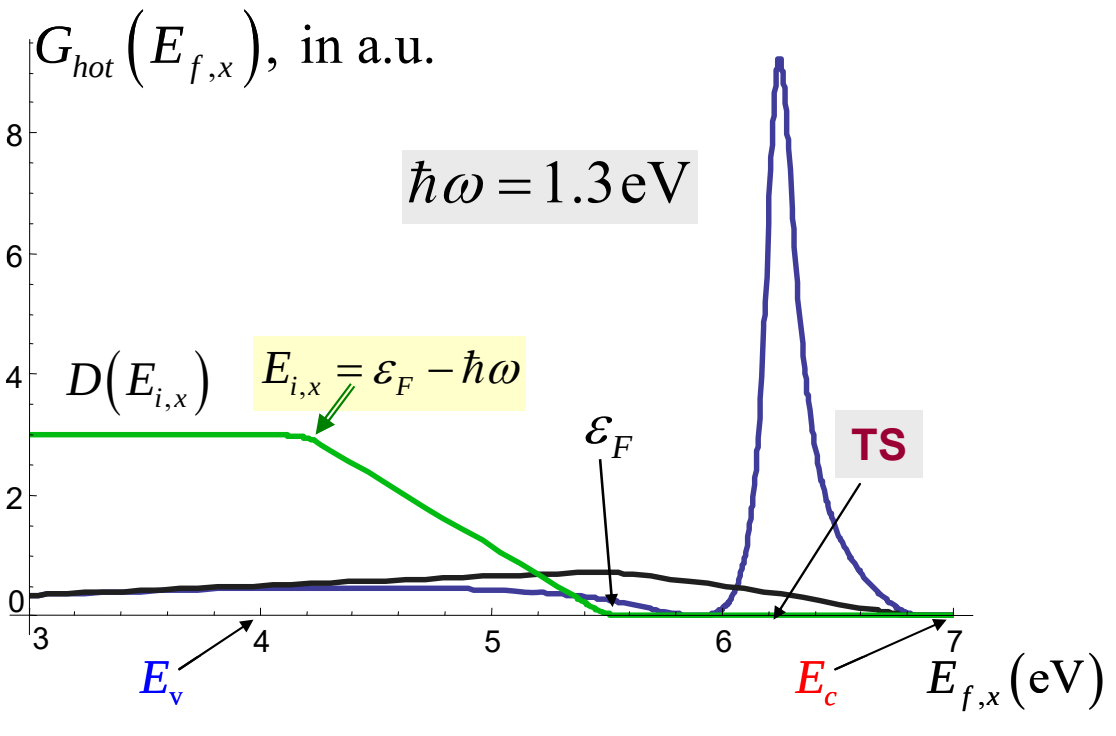

**Figure 4.** The hot electron generation rate distribution $G_{hot}(E_{f,x})$ for $\hbar\omega$=1.3eV, $d/a$ =0.4, $b/a$ =0.04, $\varepsilon_s$ =2 is shown by the blue curve. The black curve is $G_{hot}(E_{f,x})$ for the infinite barrier at the metal interface ($U_b = \infty$). The green curve illustrates the initial

electron distribution $D(E_{i,x},\hbar\omega)$ in metal, where the green arrow indicates the bend in the curve at $E_{i,x} = \varepsilon_F - \hbar\omega$.

Fig. 5 is a plot of the spectrum of the coefficient $A_{mat}$ for various sets of parameters, characterizing the metal-semiconductor interface: the barrier thickness *d*, the shift *b* and the semiconductor permittivity $\varepsilon_s$. The black curve is for the case of an infinite barrier at the interface. It is clearly seen that the coefficient $A_{mat}$ in the structure can be larger or smaller than the coefficient for the infinite barrier case, depending on the structural parameters of the interface which determine the resonant absorption at the Tamm surface state with the Fano-type shape – see Fig. 3a and Fig. 4 also. Decreasing the permittivity $\varepsilon_s$ increases the coefficient $A_{mat}$ in agreement with the fact that the transition absorption is the dominating mechanism for photon absorption as we have discussed above. In the case of infinite barriers, electrons only 'feel' the field $F_m$ in the metal since they are confined in the metal by the infinite barrier. In the structure considered, electrons can coherently penetrate into the semiconductor where the field $F_s$ is much stronger than in the metal, $|F_d| > |F_m|$. This fact obviously facilitates the LSP absorption at the interface, increasing $A_{mat}$. If an electron is in resonance with the Tamm State, its penetration into the semiconductor is enhanced (see Fig.2) whereby photon absorption is stimulated. In whole, Fig.5 demonstrates clearly that the LD and hot electron generation can be enhanced by several times in structures with TS at the interface metal-semiconductor that is attractable for applications in photocatalysis, for instance.

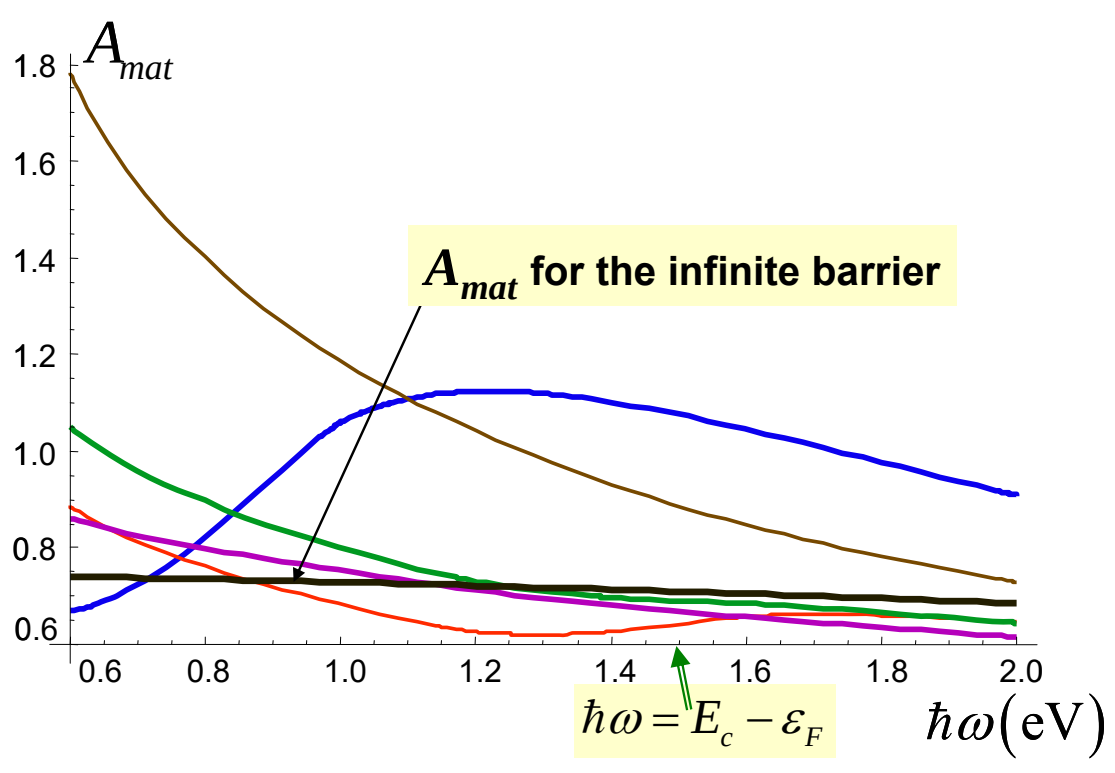

**Figure 5.** Spectrum of the coefficient $A_{mat}$: $d/a$ =0.4, $b/a$ =0.04, $\varepsilon_s$ =2 (blue); $d/a$ =0.4, $b/a$ =0, $\varepsilon_s$ =2 (red); $d/a$ =0.2, $b/a$ =0, $\varepsilon_s$ =3 (brown); $d/a$ =0.3, $b/a$ =0.01, $\varepsilon_s$ =3 (green); $d/a$ =0.3, $b/a$ =0.01, $\varepsilon_s$ =3 (violet). Black curve is for infinite barrier ($U_b = \infty$).

## 5. CONCLUSION

In summary, we have developed a model of resonant hot electron generation and enhanced Landau damping through coherent coupling of metal electrons to the Tamm surface state at the interface metal-semiconductor. The model is based on a description of the semiconductor as a 1D crystal that allows one to include, in the full band structure, the Tamm quasi-discrete energy level by natural manner. The absorption spectrum of light by electrons of the metal has a Fano-type shape due to the coherent coupling between electron states in the metal with the continuum energy spectrum and the quasi-discrete Tamm state. This coupling leads to coherent transfer of carriers between the metal and the semiconductor.

This resonant absorption is enhanced strongly by the transition absorption due to the jump of the permittivity at the interface: the real part of the metal permittivity is negative, whereas the semiconductor permittivity is positive.

Thus, the developed theory showed that the LSP absorption, modified by Tamm surface state at the interface and, more generally, by the density of surface states inside the semiconductor bandgap, affects strongly the hot electron generation and Landau damping in plasmonic nanostructures in agreement with seminal experiments on coherent electron transfer by Petek et al [27]. The hot electron generation and Landau damping both depend on the permittivity of material, surrounding metal nanoparticle, that is in accordance with earlier observations by Kreibig *et al* [7] as well as with recent experiments [16]. Thus, the LD and hot electron generation can be enhanced by several times in structures with TS at the interface metal-semiconductor.

Indeed, the present developed theory for surface states initiated absorption at the interface metal-semiconductor can be helpful for understanding of mechanisms of Chemical Interface Damping (CID) [6, 19, 20], and in particular, of the CID dependence on the permittivity of the material, surrounding plasmonic nanoparticles.

Finally, the hot electrons generated in the surface enhanced processes can be exploited in plasmonic photochemistry (photocatalysis).

## ASSOCIATED CONTENT

### Supplementary Material

1D Band structure model of semiconductors; Discrete Tamm level; Electron wave function of a quasi-discrete Tamm level

## AUTHOR INFORMATION

### Corresponding Author

*Email: uskovav@lebedev.ru

### Notes

The authors declare no competing financial interests.

## ACKNOWLEDGMENT

A.U., I.S., I.P. and N.N. are thankful to the Russian Science Foundation (Grants 20-19-00559) for support.